\documentclass[11pt]{article}
\usepackage{cite,moriond,epsfig}

\def\gtap{\mathrel{ \rlap{\raise 0.511ex \hbox{$>$}}{\lower 0.511ex
   \hbox{$\sim$}}}} 
\def\ltap{\mathrel{ \rlap{\raise 0.511ex
    \hbox{$<$}}{\lower 0.511ex \hbox{$\sim$}}}} 
\newcommand{\bea}{\begin{eqnarray}} 
\newcommand{\eea}{\end{eqnarray}}
\def\beq{\begin{equation}}
\def\enq{\end{equation}}
\def\ba{\begin{eqnarray}}
\def\ea{\end{eqnarray}}
\def\<{<\!\!}
\def\>{\!\!>}
\def\<{\langle}
\def\>{\rangle}

\begin{document}
\vspace*{4cm}
\title{
\vskip-6pt \hfill {\rm\normalsize  IPPP/08/35} \\[1ex]
\vskip-6pt \hfill {\rm\normalsize  DCPT/08/70} \\[1ex]
\vskip-12pt~\\
TESTING DARK MATTER WITH NEUTRINO DETECTORS}

\author{SERGIO PALOMARES-RUIZ}

\address{IPPP, Department of Physics, Durham University, Durham DH1
  3LE, United Kingdom}

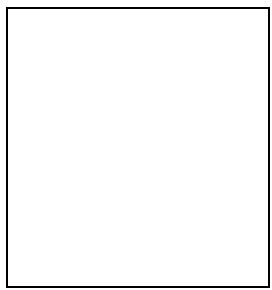
\maketitle

\abstracts{Neutrinos are the least detectable Standard Model
  particle. By making use of this fact, we consider dark matter
  annihilations and decays in the galactic halo and show how present
  and future neutrino detectors could be used to set general limits on
  the dark matter annihilation cross section and on the dark matter
  lifetime.
}

\section{Introduction}

With the next generation of neutrino experiments we will enter the era
of precision measurements in neutrino physics. As a consequence, a lot
of efforts are being dedicated to decide which are the best
experimental set-ups. However and in addition to the detailed study of
neutrino parameters, present and future neutrino detectors, thanks to
their great capabilities, might also be used for other purposes. Among
the possible synergies of these detectors, they could be used to test
some of the properties of the dark matter (DM) of the Universe. For
instance, it has been pointed out~\cite{MPP07} that by using the
spectral information of neutrinos coming from annihilations of DM
particles in the center of the Sun, some of the DM properties could be
reconstructed. In this talk however, we consider neutrinos coming from
DM annihilations or decays in our galactic halo and show how they can
be used to test some other DM properties.

We will use the fact that among the Standard Model (SM) particles,
neutrinos are the least detectable ones. Therefore, if we assume that
the only SM products from the DM annihilations (decays) are neutrinos,
a limit on their flux, conservatively and in a model-independent way,
sets an upper (lower) bound on the DM annihilation cross section
(lifetime). This is the most conservative assumption from the
detection point of view, that is, the worst possible case. Any other
channel (into at least one SM particle) would produce photons and
hence would give rise to a much more stringent limit. Let us stress
that this is not an assumption about a particular and realistic
case. On the other hand, for the reasons just stated, it is valid for
any generic model, in which DM annihilates (decays) at least into one
SM particle. Hence, the bounds so obtained are bounds on the total
annihilation cross section (lifetime) of the DM particle and not only
on its partial annihilation cross section (lifetime) due to the
annihilation (decay) channel into neutrinos.

In this talk, and following and reviewing the approach of
Refs.~\cite{BBM06,YHBA07,PP07,P07}, we consider this case and evaluate
the potential neutrino flux from DM annihilation (decay) in the whole
Milky Way, which we compare with the relevant backgrounds for
detection. In such a way, we obtain general constrains on the DM
annihilation cross section and on the DM lifetime, which are more
stringent than previous
ones~\cite{unitarity,KKT,KS,decayboundsCMB,decayboundsCMBSN}.

\section{Neutrino Fluxes from the Milky Way}

Detailed structure formation simulations show that cold DM clusters
hierarchically in halos which allows the formation of large scale
structure in the Universe to be successfully reproduced. In the case of
spherically symmetric matter density with isotropic velocity
dispersion, the simulated DM profile in the galaxies can be
parametrized via
\begin{equation}
\rho(r) = \rho_{\rm sc} \,
  \left(\frac{R_{\rm sc}}{r}\right)^\gamma \, 
  \left[\frac{1+(R_{\rm sc}/r_{\rm s})^\alpha}{1+
  (r/r_{\rm s})^\alpha}\right]^{(\beta-\gamma)/\alpha},  
\label{rhopar}
\end{equation}
where $R_{\rm sc}=8.5$~kpc is the solar radius circle, $\rho_{\rm sc}$
is the DM density at $R_{\rm sc}$, $r_{\rm s}$ is the scale radius,
$\gamma$ is the inner cusp index, $\beta$ is the slope as $r
\rightarrow \infty$ and $\alpha$ determines the exact shape of the
profile in regions around $r_{\rm s}$. Commonly used profiles
~\cite{Moore,NFW,Kravtsov} (see also Ref.~\cite{DMprofiles}) tend to
agree at large scales, although they differ considerably in the inner
part of the galaxy. 
 
The differential neutrino plus antineutrino flux per flavor
from DM annihilation or decay in a cone of half-angle $\psi$ around
the galactic center, covering a field of view $\Delta\Omega = 2 \, \pi
\, \left(1 - \cos\psi \right)$, is given by
\begin{equation}
\frac{ d \Phi}{d E_\nu} = \frac{\Delta\Omega}{4\, \pi} \, {\cal P}_k
 (E_\nu, m_\chi) \, R_{\rm sc} \,
 \rho_0^k \, {\cal
 J}_{\Delta\Omega, k} \, , 
\label{dkflux}
\end{equation}
where $m_\chi$ is the DM mass, $\rho_0 = $ 0.3~GeV~cm$^{-3}$ is a
normalizing DM density, which is equal to the commonly quoted DM
density at $R_{\rm sc}$, and ${\cal J}_{\Delta\Omega, k}$ is the
average in the field of view (around the galactic center) of the line
of sight integration of the DM density (for decays, $k=1$) or of its
square (for annihilations, $k=2$), which is given by 
\begin{equation}
{\cal J}_{\Delta\Omega, k}  = \frac{2 \, \pi}{\Delta\Omega} \,
 \frac{1}{R_{\rm sc} \, \rho_0^k}  \, \int_{\cos \psi}^1 \,
 \int_{0}^{l_{\rm max}} \,  \rho (r)^k \, dl \, d(\cos \psi'), 
\label{Javg}
\end{equation}
where $r = \sqrt{R^2_{\rm sc} -  2 l R_{\rm sc} \cos \psi' + l^2}$
and $l_{\rm max} = \sqrt{(R_{\rm halo}^2 - \sin^2 \psi R^2_{\rm sc})} +
R_{\rm sc} \cos \psi$. The contribution at large scales is negligible
and thus, this integral barely depends on the size of the
halo for $R_{\rm halo} \gtap$~few tens of kpc.

The factor ${\cal P}_k$ embeds all the dependences on the particle
physics model and it reads 
\begin{equation}
{\cal P}_1 = \frac{1}{3} \, \frac{dN_1}{dE_\nu} \, \frac{1}{m_\chi
  \tau_\chi} \hspace{5mm} {\rm for \, \, decays \, \, and} \hspace{5mm}
{\cal P}_2 = \frac{1}{3} \, \frac{dN_2}{dE_\nu} \, \frac{\langle
  \sigma_{\rm A} v \rangle}{2 \, m_\chi^2} \hspace{5mm} {\rm for \, \,
  annihilations} \, ,
\end{equation}
where the neutrino plus antineutrino spectrum per flavor is given by
\begin{equation}
\frac{dN_1}{dE_\nu} = 2 \, \delta (E_\nu - \frac{m_\chi}{2})
     \hspace{5mm} {\rm for \, \, decays \, \, and} \hspace{5mm} 
\frac{dN_2}{dE_\nu} = 2 \, \delta (E_\nu - m_\chi) \hspace{5mm} {\rm
  for \, \, annihilations} \, ,
\end{equation}
and the factor of 1/3 comes from the assumption that the
annihilation or decay branching ratio is the same for the three
neutrino flavors. Let us note that this is not a very restrictive
assumption, for even even when only one flavor is predominantly
produced, there is a guaranteed flux of neutrinos in all flavors
thanks to the averaged neutrino oscillations between the source and the
detector. Hence, although different initial flavor ratios would give 
rise to different flavor ratios at detection, the small differences
affect little our results and for simplicity herein we consider flavor
democracy.

\subsection{Annihilations versus Decays: DM Halo Uncertainties}

As mentioned above, while DM profiles tend to agree at large scales,
uncertainties are still present for the inner region of the galaxy. In
the two cases considered (annihilations and decay), the overall
normalization of the flux is affected by the value of ${\cal
  J}_{\Delta\Omega, k}$. However, in the case of DM annihilations,
it scales as $\rho^2$, whereas for DM decays, it scales as
$\rho$. Our lack of knowledge of the halo profile is hence much more
important for the neutrino flux from DM annihilations. For the three
profiles considered here~\cite{Moore,NFW,Kravtsov}, astrophysical
uncertainties can induce errors of up to a factor of 6 for the case of 
DM decays~\cite{P07}, but they can be as large as a factor of
$\sim$~100 for DM annihilations~\cite{YHBA07,PP07}. In addition, if
the DM mass is not known, DM annihilation and DM decay in the halo
might have the same signatures. However, due to the fact that the
dependence on the DM halo density is different for each case, in case
of a positive signal, directional information would be crucial to
distinguish between these two possibilities.

For concreteness, in what follows we present results using the
Navarro, Frenk and White (NFW) simulation~\cite{NFW} as our canonical
profile.

\section{Neutrino Bounds}

In order to obtain the constraints on the DM annihilation cross
section and DM lifetime we assume that DM
annihilates~\cite{BBM06,YHBA07,PP07} or decays~\cite{P07} only into
neutrinos. If DM annihilates or decays into SM particles, neutrinos
(and antineutrinos) are the least detectable ones. Any other possible
annihilation or decay mode would produce gamma rays, which are much
easier to detect, and would allow to set a much stronger (and
model-dependent) bound. Thus, the most conservative
approach~\cite{BBM06,YHBA07,PP07,P07} is to assume that only neutrinos
are produced in DM annihilations or decays. Even in this conservative
case, it has been shown that stringent limits can be obtained by
comparing the expected time-integrated annihilation signal of all
galactic halos~\cite{BBM06} and the signal from
annihilations~\cite{YHBA07,PP07} or decays~\cite{P07} in the Milky Way
Halo with the background at these energies.

\subsection{The Atmospheric Neutrino Background}

For $E_\nu \gtap$~100~MeV, the main source of background for a
possible neutrino signal from DM annihilations or decays is the flux
of atmospheric neutrinos, which is well known up to energies of
$\sim$~100~TeV. Thus, in order to obtain a bound on the DM annihilation
cross section and lifetime we need to compare these two fluxes, and in
particular we consider the $\nu_\mu + \overline{\nu}_\mu$ spectra
calculated with FLUKA~\cite{FLUKAhigh}. 

In this energy range, we will follow the approach of
Ref.~\cite{YHBA07}. By assuming that the only resultant products of DM
annihilation (decay) are neutrino-antineutrino pairs, we first obtain
a general bound by comparing the ($\nu_\mu + \overline{\nu}_\mu$)
neutrino flux from DM annihilation (decays) in the halo with the
corresponding atmospheric neutrino flux for $E_\nu
\sim$~100~MeV--100~TeV in an energy bin of width $\Delta \log_{10}
E_\nu = 0.3$ around $E_\nu = m_\chi$ ($E_\nu = m_\chi/2$). For each
value of $m_\chi$, the limit on $\langle \sigma_{\rm A} v \rangle$
($\tau_\chi$) is obtained by setting its value so that the neutrino
flux from DM annihilations (decays) in the Milky Way equals the
atmospheric neutrino spectrum integrated in the chosen energy bin. The
reason for choosing this energy bin is mainly that the neutrino signal
is sharply peaked around a neutrino energy equal to the DM mass (half
of the DM mass) and this choice is within the experimental limits of
neutrino detectors. 

\begin{figure}
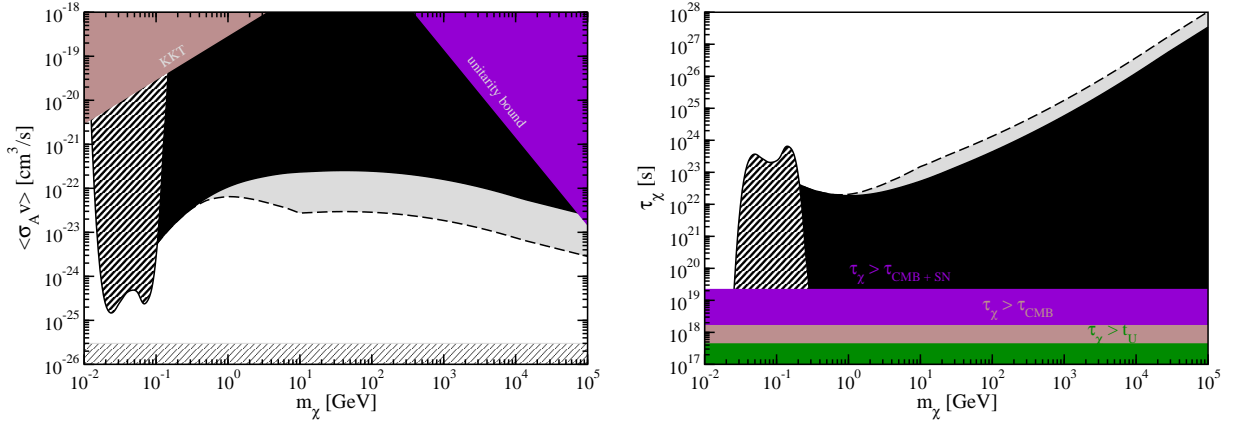

\epsfig{figure=DMcsboundMoriond08.eps,height=2.2in}
\hspace{3mm}
\epsfig{figure=DMlifetimeboundMoriond08.eps,height=2.2in}
\caption{Bounds on the total DM annihilation cross section (left
  panel) and DM lifetime (right panel) for a wide range of DM masses
  obtained using different approaches: full-sky signal (dark area), 
  angular signal (light area) and 90\% CL limit using SK data at low
  energies~\protect\cite{SKSN} (hatched area). Results are obtained
  for a NFW profile. Other general bounds are also shown. Right panel:
  the unitarity bound~\protect\cite{unitarity}, the limit above which
  the cusps of the DM halos are too flat (KKT)~\protect\cite{KKT}
  and the natural scale for thermal relics. Left panel: bounds from
  Cosmic Microwave Background
  observations~\protect\cite{decayboundsCMB} and Cosmic Microwave
  Background plus Supernovae 
  data~\protect\cite{decayboundsCMBSN} (both at 2$\sigma$ confidence
  level) and the line $\tau_\chi = t_U$, with $t_U \simeq 4 \times
  10^{17}$~s the age of the Universe. Adapted from
  Refs.~\protect\cite{PP07,P07}.
}
\label{DMbound}
\end{figure}

The most conservative bound is obtained by using the full-sky signal,
and this is shown in both panels of Fig.~\ref{DMbound} where the dark
areas represent the excluded regions. However, a better limit can be
obtained by using angular information. This is mainly limited by the
kinematics of the interaction. In general, neutrino detectors are only
able to detect the produced lepton and its relative direction with
respect to the incoming neutrino depends on the neutrino energy as
$\Delta\theta \sim 30^o \times \sqrt{{\rm GeV}/E_\nu}$. As in
Ref.~\cite{YHBA07} and being conservative, we consider a field of view
with a half-angle cone of $30^o$ ($30^o \times \sqrt{10 \, {\rm
    GeV}/E_\nu}$) for neutrinos with energies above (below)
10~GeV. This limit is shown in both panels of Fig.~\ref{DMbound} by
the dashed lines (light areas), which improves upon the previous case
by a factor of a few for $E_\nu >$~5~GeV.

\subsection{MeV Dark Matter}

As we have just described, it is expected that a more detailed
analysis, making a more careful use of the directional as well as
energy information for a given detector, will improve these
results. Note for instance that for energies $\sim$~1-100~GeV neutrino
oscillations would give rise to a zenith-dependent background, whereas
we expect a nearly flat background for other energies for which
oscillations do not take place. We now show how a more careful
treatment of the energy resolution and backgrounds can substantially
improve these limits~\cite{PP07,P07}.

Here we describe the analysis followed in Refs.~\cite{PP07,P07} to set
neutrino constraints on the DM total annihilation cross section and DM
lifetime in the energy range 15~MeV~$ \ltap E_\nu \ltap 130$~MeV. In
this energy range the best data comes from the search for the diffuse
supernova background by the Super-Kamiokande (SK) detector which has
looked at positrons (via the inverse beta-decay reaction,
$\overline{\nu}_e + p \rightarrow e^+ + n$) in the energy interval 
18~MeV--82~MeV~\cite{SKSN}. As for these energies there is no
direction information, we consider the full-sky $\overline{\nu}_e$
signal. In this search, the two main sources of
background are the atmospheric $\nu_e$ and $\overline{\nu}_e$ flux
and the Michel electrons and positrons from the decays of
sub-threshold muons. Below 18~MeV, muon-induced spallation products
are the dominant background, and below $\sim$~10~MeV, the signal
would be buried below the reactor antineutrino background.

Although for $E_\nu \ltap$~80~MeV the dominant interaction is the
inverse beta-decay reaction (with free protons), the interactions of
neutrinos (and antineutrinos) with the oxygen nuclei contribute
significantly and must be considered. For our analysis we have
included both the interactions of $\overline{\nu}_e$ with free protons
and the interactions of $\nu_e$ and $\overline{\nu}_e$ with bound
nucleons, by considering, in the latter case, a relativistic Fermi gas
model~\cite{SM72} with a Fermi surface momentum of 225~MeV and a
binding energy of 27~MeV. We then compare the shape of the background
spectrum to that of the signal by performing a $\chi^2$ analysis,
analogous to that of the SK collaboration~\cite{SKSN}. In this way, we 
can extract the limits on the DM annihilation cross section and DM 
lifetime~\cite{PP07,P07}. Hence, we consider the sixteen 4-MeV bins in
which the data were divided and define the following $\chi^2$
function~\cite{SKSN}
\begin{equation}
\chi^2 = \sum_{l=1}^{16} \, \frac{\left[(\alpha \cdot A_l) + (\beta
    \cdot B_l) + (\gamma \cdot C_l) - N_l \right]^2}{\sigma_{stat}^2 +
    \sigma_{sys}^2} ~,
\end{equation}  
where the sum $l$ is over all energy bins, $N_l$ is the number of
events in the $l$th bin, and $A_l$, $B_l$ and $C_l$ are the fractions
of the DM annihilation or decay signal, Michel electron (positron) and
atmospheric $\nu_e$ and $\overline{\nu}_e$ spectra that are in the
$l$th bin, respectively. The fractions $A_l$ are calculated taking
into account the energy resolution of SK, interactions with free and
bound protons and the correct differential cross
sections~\cite{PP07}. The fractions $B_l$ are calculated taking into 
consideration that in water 18.4\% of the $\mu^-$ produced below
\v{C}erenkov threshold ($p_\mu < 120$~MeV) get trapped and enter a
K-shell orbit around the oxygen nucleus and thus, the electron
spectrum from the decay is slightly distorted with respect to the
well-known Michel spectrum~\cite{HVRA74}. In the calculation of the
fractions $B_l$ and $C_l$ we have used the low energy atmospheric
neutrino flux calculation with FLUKA~\cite{FLUKAlow}. Note that, in a
two-neutrino approximation and for energies below $\sim$300~MeV (where
most of the background comes from), half of the $\nu_\mu$ have
oscillated to $\nu_\tau$, whereas $\nu_e$ remain unoscillated.
Although this approximation is not appropriate, in principle, to
calculate the low energy atmospheric neutrino background, however, for
practical purposes, it introduces very small
corrections~\cite{FLMM04}. Thus, in order to calculate $B_l$ and $C_l$ 
we use the two-neutrino approximation. The fitting parameters in the
$\chi^2$-function are $\alpha$, $\beta$ and $\gamma$, which represent
the total number of each type of event. For the systematic error we
take $\sigma_{sys} = 6 \%$ for all energy bins~\cite{SKSN}. 

In absence of a DM signal, a 90\% confidence level (C.L.) limit can be
set on $\alpha$ for each value of the DM mass. The limiting
$\alpha_{90}$ is defined as
\begin{equation}
\int_{0}^{\alpha_{90}} P(\alpha) \, d\alpha = 0.9 \, ,
\end{equation}
where~\footnote{Note that there is an error in Eq.(8) of
  Ref.~\cite{PP07}. Nevertheless, this implies very small corrections
  to the results presented. I thank O.~L.~G.~Peres for pointing this
  out.} $P(\alpha) = K \cdot e^{-\chi_{\alpha}^2/2}$ is the relative
probability and $\chi_{\alpha}^2$ is the minimum $\chi^2$ for each
$\alpha$. The normalizing constant $K$ is such that $\int_{0}^{\infty}
P(\alpha) \, d\alpha = 1$. It is straightforward to translate the limit
on $\alpha$ into limits of the total DM annihilation cross section and
DM lifetime and these 90\% CL bounds are shown in both panels of 
Fig.~\ref{DMbound} by the hatched areas and they clearly
improve (and extend to lower masses) by about an order of magnitude 
upon the general and very conservative bound obtained with the simple
analysis described above for higher energies.

\section{Conclusions}

In this talk we have shown how neutrino detectors can also be used to
test some of the DM properties and have obtained general bounds on the
DM annihilation cross section and DM lifetime, which greatly improve
over previous
limits~\cite{unitarity,KKT,KS,decayboundsCMB,decayboundsCMBSN}. In
order to do so, we have assumed that the only SM products from DM 
annihilations or decays are neutrinos, which are the least detectable
particles of the SM. By making this assumption we have obtained 
conservative but model-independent bounds. In a simple way and for
energies between $\sim$~100~MeV and $\sim$~100~TeV, we have considered
the potential signal from DM annihilations or decays in the Milky Way
and have compared it to the atmospheric neutrino background. The
general bounds are obtained by considering this potential signal and 
imposing that it has to be at most equal to the background in a given
energy interval. We have also shown how this crude, but already very
stringent limit, can be substantially improved by more detailed
analysis which make careful use of the angular and energy resolution
of the detectors, as well as of backgrounds. In this way, we have
obtained~\cite{PP07,P07} the 90\% CL bounds on the DM annihilation
cross section and DM lifetime for $m_\chi \ltap$~200~MeV, which is
about an order of magnitude more stringent.

\section*{Acknowledgments}
The author is partially supported by the Spanish Grant FPA2005-01678
of the MCT.

\end{document}